\documentclass[a4paper,11pt]{article}

\usepackage{jinstpub}

\usepackage{graphicx}
\usepackage{amssymb}
\usepackage{amsmath}
\usepackage{amstext}

\usepackage{lineno}

\usepackage{xspace}
\usepackage{xcolor}
\usepackage[normalem]{ulem}

\newcommand{\ttbar}{\ensuremath{t\bar{t}}\xspace}

\newcommand{\Matriplex}{\textsc{Matriplex}\xspace}
\newcommand{\mplex}{\textsc{Matriplex}\xspace}
\newcommand{\mkFit}{\textsc{mkFit}\xspace}
\newcommand{\cpp}{C\texttt{++}\xspace}

\newcommand{\Intel}{Intel\textregistered\xspace}
\newcommand{\Xeon}{Xeon\textregistered\xspace}
\newcommand{\XeonPhi}{Xeon Phi\texttrademark\xspace}

\newcommand{\unit}[1]{\ensuremath{\text{\,#1}}\xspace}           

\title{Speeding up Particle Track Reconstruction using a Parallel Kalman Filter Algorithm}

\author[a]{Steven Lantz,}
\author[a]{Kevin McDermott,}
\author[a]{Michael Reid,}
\author[a]{Daniel Riley,}
\author[a]{Peter Wittich,}
\author[b]{Sophie Berkman,}
\author[b]{Giuseppe Cerati,}
\author[b]{Matti Kortelainen,}
\author[b]{Allison Reinsvold Hall,}
\author[c]{Peter Elmer,}
\author[c]{Bei Wang,}
\author[d]{Leonardo Giannini,}
\author[d]{Vyacheslav Krutelyov,}
\author[d]{Mario Masciovecchio,}
\author[d]{Matev\v{z} Tadel,}
\author[d]{Frank W\"{u}rthwein,}
\author[d]{Avraham Yagil,}
\author[e]{Brian Gravelle,}
\author[e]{Boyana Norris.}

\affiliation[a]{Cornell University, Ithaca, NY, USA 14853}
\affiliation[b]{Fermi National Accelerator Laboratory, Batavia, IL, USA 60510}
\affiliation[c]{Princeton University, Princeton, NJ, USA 08544}
\affiliation[d]{UC San Diego, La Jolla, CA, USA 92093}
\affiliation[e]{University of Oregon, Eugene, OR, USA 97403}

\emailAdd{mic-trk-rd@cern.ch}

\abstract{One of the most computationally challenging problems expected for the High-Luminosity Large Hadron Collider (HL-LHC) is determining the trajectory of charged particles during event reconstruction. Algorithms used at the LHC today rely on Kalman filtering, which builds physical trajectories incrementally while incorporating material effects and error estimation. Recognizing the need for faster computational throughput, we have adapted Kalman-filter-based methods for highly parallel, many-core SIMD architectures that are now prevalent in high-performance hardware. In this paper, we discuss the design and performance of the improved tracking algorithm, referred to as \mkFit. A key piece of the algorithm is the \Matriplex library, containing dedicated code to optimally vectorize operations on small matrices. The physics performance of the \mkFit algorithm is comparable to the nominal CMS tracking algorithm when reconstructing tracks from simulated proton-proton collisions within the CMS detector. We study the scaling of the algorithm as a function of the parallel resources utilized and find large speedups both from vectorization and multi-threading. \mkFit achieves a speedup of a factor of 6 compared to the nominal algorithm when run in a single-threaded application within the CMS software framework.}


\begin{document}
\maketitle
\flushbottom


\section{Introduction}

\subsection{Physics Motivation and Goals}

The experimental challenges of High Energy Physics (HEP) experiments such as the Compact Muon Solenoid~\cite{Chatrchyan:2008aa} (CMS) at the CERN Large Hadron Collider~\cite{Evans:2008zzb} (LHC) are vast. The rate at which interesting processes (signal) are produced in the beam collisions is many orders of magnitude lower than the most common and well-known processes (background) occurring in proton interactions. For this reason, experiments need high proton collision rates (``instantaneous luminosity"). For a fixed proton bunch colliding frequency (40~MHz at the LHC), the instantaneous luminosity is increased by squeezing the beams so that they have larger proton density at the beam colliding point; this leads to a larger number of simultaneous proton interactions for each bunch crossing, commonly referred to as ``pileup" (PU).  In the upcoming upgrade of the accelerator from the LHC to the High-Luminosity LHC~\cite{ApollinariG.:2017ojx} (HL-LHC), the instantaneous luminosity will increase by a factor of 5 with respect to the current LHC operations, leading to PU values as high as 200.

Experiments are read out at the beam collision rate, and each readout defines an ``event". The resulting amount of data recorded by the experiments is so large that it cannot be entirely stored and processed. A ``trigger" system filters interesting events in real time, where a first selection is applied at hardware level (L1 trigger) followed by a software-based selection (High Level Trigger, HLT, or ``online" processing)~\cite{CMSTrigger}. Events selected by the trigger are saved and subsequently processed for analysis (``offline" processing).

A fundamental component of both online and offline processing is event reconstruction, i.e., the process of converting raw signals from the many detector elements into higher-level physics observables. With increasing PU, reconstruction algorithms are facing increasing complexity in the data, mostly due to the higher detector occupancy; this implies that for combinatorial algorithms the time needed to process one event increases dramatically. Higher PU values also reduce the precision with which an event can be reconstructed, typically leading to increased backgrounds.

Computing resources available to the experiments define a time budget for reconstruction processing. Requirements are particularly strict at the HLT, where data is processed in a dedicated computing farm and a yes/no decision needs to be made at a rate of 100 kHz. The rate will increase to 750 kHz during the HL-LHC. Reconstruction is both the driver of the per-event processing time and of the selection quality: more advanced reconstruction algorithms allow interesting physics events to be selected with more surgical precision. Reconstruction time also limits offline processing, in terms of the size of data and simulation samples produced in a given time frame; 
clearly the capability to reprocess the data to take advantage of improved calibrations, as well as being able to generate larger simulated samples for more precise signal and background predictions, has a direct impact on the physics output of the experiments. 

The work presented in this paper focuses on a particular component of the event reconstruction: track reconstruction or tracking. Tracking is the reconstruction of charged particle trajectories from the energy deposits (``hits") they produce in the detector. It is by far the most time-consuming step in the whole reconstruction process, both online and offline, and, due to its combinatorial nature, its processing time diverges at large PU~\cite{Cerati:2015vna}. By the start of the HL-LHC, the expected CPU needs of the LHC experiments are projected to exceed the available resources by a factor of 4~\cite{IRISHEP}. 

An option to keep the tracking time under control is to limit the reconstruction to high-momentum particles or ``regions of interest'' as seeded by energy deposits in other subdetectors. Similarly, one could consider not reconstructing displaced tracks, or those originating from PU vertices. However, these reductions involve real physics trade-offs.  For instance, not reconstructing low-momentum tracks has implications for being able to detect leptons from soft SUSY particles~\cite{Khachatryan:2015pot}; tracks from PU interactions are used to measure, and then mitigate, the effect of PU on an event-by-event basis~\cite{Bertolini:2014bba}; displaced tracks are critical to detect $b$ quark decays~\cite{Sirunyan:2017ezt}, for tau lepton tagging~\cite{Sirunyan:2018pgf} and to reconstruct photon conversions~\cite{Khachatryan:2010pw}, which are needed for maximal acceptance in the $H\to\gamma\gamma$ decay channel~\cite{Khachatryan:2014ira}. So while it might be possible to speed up tracking by neglecting certain kinds of tracks, this leads to unacceptable physics compromises. 

\subsection{Moore's Law Transformed: Track Reconstruction Opportunities}
\label{sec:moore-trk}
Solutions for speeding up tracking with no physics compromises come from recent developments in computing architectures. Moore's Law states that chip transistor counts increase exponentially over time for the same cost.  Over four decades, these increases in transistor counts turned into exponential gains in the performance of software applications like those used in particle physics.  Around 2005, the computing processor market reached an epochal turning point: power density limitations in chips ended this trend, so that serial applications no longer immediately run exponentially faster on subsequent generations of processors.  This is true even though the underlying transistor count continues to increase per Moore's Law. What changed is that the extra transistors now provide processors with more and more parallel elements running at roughly the same speed as in prior generations.

What are these power-efficient ``parallel elements''? In microprocessors, they take two main forms: multiple cores, each of which is a CPU in its own right; and vector processing units, which add ``Single Instruction, Multiple Data'' (SIMD) execution capability to these cores. Thus for CPUs in the present phase of Moore's Law, the extra parallelism is resulting in higher per-processor core counts and wider vector processing units. Another variant occurs in General Purpose Graphics Processing Units (GPGPUs, or simply GPUs), where
the parallel elements take the form of thousands of relatively simple stream processors; these are likewise growing in number over time.

The evolution also appears to be toward heterogeneous mixes of these solutions. For instance, the CPU-based nodes of a computing cluster may be augmented with GPU accelerator cards. Such a mix presents a challenging target to application programmers. However, it turns out that modern CPUs and GPUs both favor the same kinds of parallelism in applications: namely, multiple threads or streams of execution, each of which is executed in SIMD fashion. Thus, while particular implementations may vary, there is a clear indication of the general evolutionary path that high-performance software will need to take, in light of the commonly available hardware.


\section{Overview and Objectives}

This paper describes the design, implementation, and performance of a tracking algorithm known as \mkFit. 
The code developed in this project 
allows for a paradigm shift by switching tracking software from being sequential to becoming both multithreaded and vectorized. 
The \mkFit algorithm is a new implementation of the traditional Kalman filter (KF) approach for charged particle tracking~\cite{Fruhwirth:1987fm}. KF algorithms have been used effectively in many particle collider experiments such as CMS and ATLAS, but are typically difficult to vectorize due to the many branch points required to explore different track candidates.
The project mostly targets multicore CPU architectures such as the \Intel \Xeon and \Intel \XeonPhi processors and coprocessors, but the general SIMD and parallelization strategies employed by \mkFit would allow the algorithm to run efficiently on other multicore CPUs as well. 
Initial explorations into a GPU implementation have also been performed, but will not be discussed extensively in this article.

In order to achieve the granularity needed to perform pattern recognition in very dense environments, modern HEP tracking detectors consist of millions of sensors arranged in multiple layers around the interaction region.
Tracking traditionally proceeds in three main stages: seeding, building, and fitting. Seeding provides the initial estimate of the track parameters based on a few hits in a subset of the innermost detector layers; since seeding typically does not employ KF techniques, it is not part of this work. 
Building then collects additional hits in other detector layers to form a complete track.
After hits have been assigned to each track, a final fit is performed to provide the best estimate of the track parameters. Track building is the most time-consuming step of HEP event reconstruction and is the focus of this project. 

The initial objective of the \mkFit project is to demonstrate substantial speedups and similar physics performance compared to the offline track building algorithm used by CMS and implemented in CMSSW, the software framework for the experiment~\cite{Jones:2006cmssw}.
The CMS track building software uses an iterative tracking approach, where each iteration removes hits associated with found tracks, thereby reducing the combinatorial complexity for subsequent iterations. The \mkFit algorithm is currently applied to the first iteration of the offline CMS track building, which is responsible for building the majority of high-quality, prompt tracks. With appropriate bookkeeping and re-tuning of the tracking parameters, \mkFit could be adapted for the other CMS tracking iterations as well. Additionally, work is ongoing to test the performance of the \mkFit algorithm in the context of the CMS HLT.

Section~\ref{sec:kalman_intro} describes the essential components of a Kalman-filter-based tracking algorithm. In Section~\ref{sec:algorithm}, we describe the algorithm's implementation and parallelization strategy.
The physics and computational performance of the algorithm are demonstrated in Section~\ref{sec:Results}, and the outlook and future plans for the project are discussed in Section~\ref{sec:outlook}.

\section{Kalman Filter Tracking}
\label{sec:kalman_intro}

Kalman filtering~\cite{Kalman1960} is a technique that determines the internal state of a linear dynamic system by recursively processing discrete measurements where random perturbations are present both in the measurements and in the system itself. While its most common application is in the navigation of aircraft and ships, it is also applied to a much broader variety of domains, including particle tracking in HEP experiments~\cite{Fruhwirth:1987fm}. 

In a HEP tracking detector, every sensor detects and reads out the ionization charge deposited by impinging particles. Such detectors require electronics, cables, cooling systems, and support structures for a total material budget frequently exceeding one radiation length. Furthermore, the entire detector is immersed in a strong magnetic field (usually homogeneous) to make the charged particle trajectories roughly helical, so that each particle's momentum can be inferred from its helix curvature. KF-based tracking algorithms are widely used because they naturally incorporate estimates of material effects (deviations due to multiple scattering, energy loss) while fitting the trajectory of a given particle to a helix.

Other algorithms originating in the image processing community are amenable to parallelization, and they have been explored by different groups. These include Hough Transforms~\cite{Halyo:2013gja} and Cellular Automata~\cite{Pantaleo:2293435}, among others. However, they are not the main algorithms in use at the LHC today for track building, while KF algorithms have proven to be robust and perform well in the difficult experimental environment of the LHC~\cite{Chatrchyan:2014fea,Aaboud:2017all}. Rather than abandon the collective understanding of how KF algorithms perform in the HEP context, we wish to extend this well-known tool by designing an optimal implementation on highly parallel architectures.

\begin{figure}[h]
\centering\includegraphics[width=0.95\linewidth]{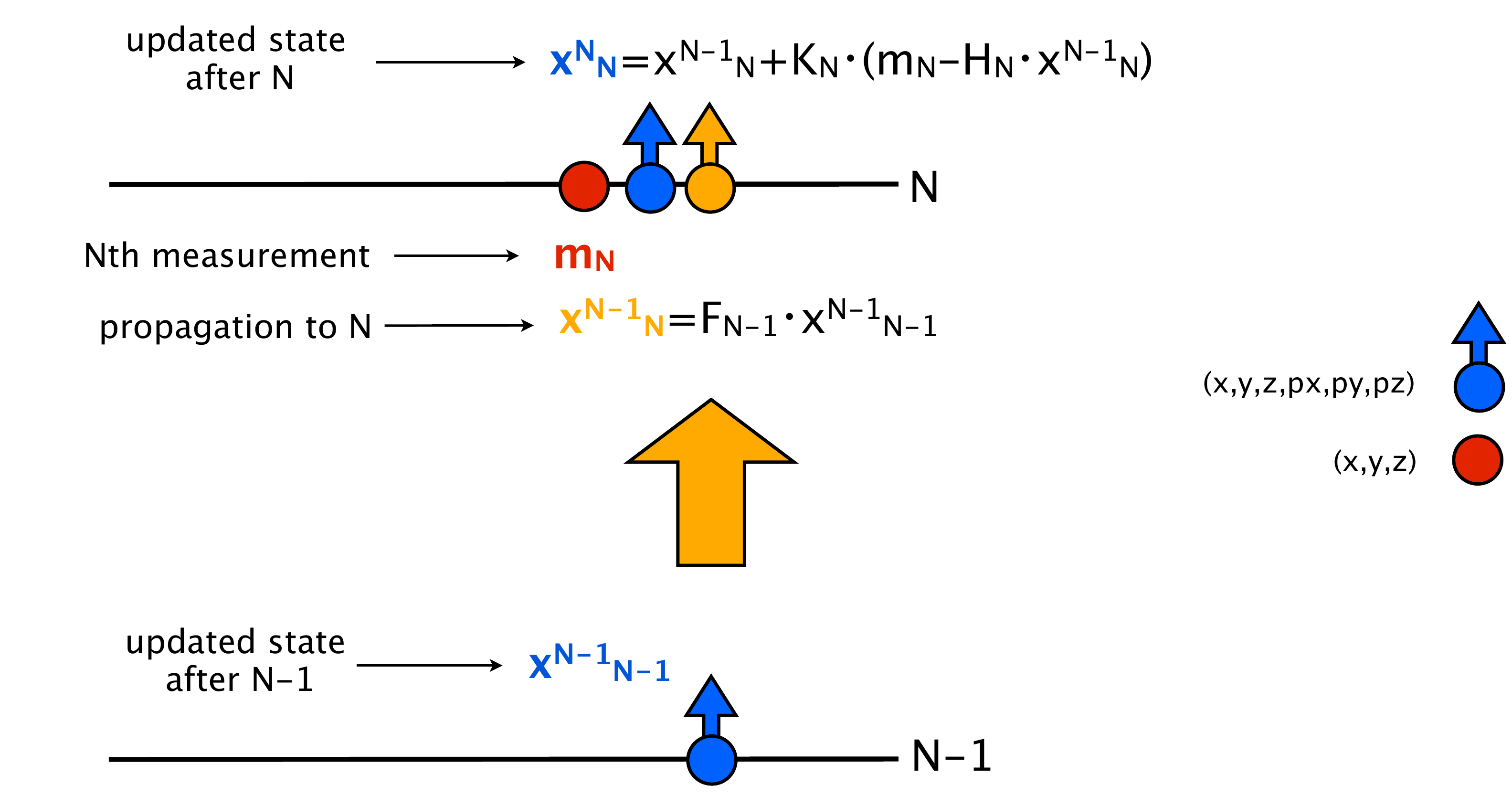}
\caption{Representation of the logic unit of the Kalman filter for charged particle tracking, where the notation follows conventions in~\cite{Fruhwirth:1987fm}. The track state (x) is propagated from layer N-1 to layer N (using a linear function F for simplicity), and then updated with the measurement on layer N ($\mathrm{m_N}$) using the ``Measurement Matrix" H and the ``Kalman Gain" matrix K.}
\label{fig:kalman-unit}
\end{figure}

The core logic unit of the KF algorithm is shown in Figure~\ref{fig:kalman-unit}. This unit forms the basis for both the track building and the track fitting steps.
The unit begins with the propagation of the \emph{track state}, defined as the track parameters and corresponding uncertainties (covariance matrix), from one layer to the next. At the new layer, a $\chi^2$ is computed between the propagated track state and the hit measurement on that layer. Finally, the track state is updated using the information from the hit measurement. 

Fitting simply consists of the iterative application of the KF logic unit over a defined set of hits associated 
to a track. Therefore, to first order, its speed scales linearly with the number of tracks in each event.
Building, instead, needs to identify at each layer which hits are compatible with a given track candidate, if any.
Starting from a seed, for each compatible hit a new candidate is created and propagated to the next layer.
Track candidates originating from the same seed are sorted according to a score function (depending on the number of found hits, the number of layers crossed without a corresponding hit being found, and the total $\chi^2$) so that the best ones are retained at each layer. At the end of the process, the candidate with the best score is selected. For a more detailed description of the algorithm logic, see~\cite{Chatrchyan:2014fea}.
Due to its combinatorial nature, building is by far the most time-consuming step of tracking and it scales worse than linearly with increasing occupancy. For these reasons, track building has been the primary focus of the \mkFit project.

\section{Algorithm Design and Implementation}
\label{sec:algorithm}

\subsection{Challenges and Drivers For Algorithm Design}

The promise of highly parallel architectures comes with significant trade-offs, so that algorithms need to be
specifically designed to run efficiently on them. First, these architectures commonly feature two types of parallelism: vectorization and multithreading. Vector or SIMD operations perform a single instruction on multiple data at the same time, in lockstep;
therefore, branching points in the algorithm may lead to significant performance degradation. 
Multithread parallelism, on the other hand, involves performing different instructions on different data at the same time; in this
case, the factors limiting performance are load balancing and synchronization between threads.
In addition, threads may be in contention for data from memory, since a processor's memory bandwidth and caches are limited resources. Therefore, algorithms are penalized unless their data structures are refactored to minimize memory usage and to favor regular access patterns such as ``structures of arrays''. 
In a nutshell, an efficient parallel algorithm must have its whole flow engineered toward ensuring that all the parallel processing units are constantly occupied with floating-point operations.

The above factors prevent the straightforward parallelization of an existing piece of code of sufficient complexity. KF tracking in particular faces a number of challenges.
Due to its combinatorial nature, branching is at the very heart of KF track building: multiple combinations of hit patterns must be explored and compared in order to find the best one. Indeed, every hit constitutes a possible branching point: if a hit is deemed compatible with the track candidate under evaluation, a new logical path is created.
Following the branches in a way that is consistent with the SIMD computing model is a significant challenge.
Furthermore, as the track candidates are built, they typically probe and select different numbers of hits, since the detector occupancy is not uniformly distributed on a per-event basis. This leads to the challenge of parallel work imbalance, at both vector and  thread levels. Finally, the combinatorial search takes place among O(100k) hits in the detector, organized in O(10k) hits per layer, and yet a track candidate needs to probe only O(1-10) hits in the vicinity of its crossing point on a given layer. The compatibility test for these few hits requires just a small number of floating point operations. This means the KF algorithm presents extra challenges in terms of data locality and in terms of arithmetic intensity (floating-point operations per memory access). 

To realize any of the potential performance gains, the problem needs to be rearranged to optimize resource usage and to fit into the parallel hardware environment. 
While the algorithm implementation will be described in detail in the next sections, it is useful to list up front the guiding principles we rely upon in its design.
First, in order to most efficiently exploit vectorization, we factorize the algorithm so that branching points are confined in specific non-vectorized functions, while the core of the KF logic unit is expressed in terms of SIMD data structures and computations.
Second, in order to utilize as many cores as possible while at the same time minimizing the load imbalance across threads, we employ a thread scheduling strategy at multiple levels, with large parallel regions corresponding to fixed boundaries in terms of detector or data partitions and with a more fine-grained parallelism at track level.
Finally, to achieve efficient usage of the memory resources, we reduce as much as possible the data structures, while organizing the data in partitions for fast access, representing the detector geometry using parametric descriptions rather than large lookup tables, and minimizing memory movements especially during SIMD processing steps.

\subsection{Parallelization Strategy}

Collision events inside a particle detector are naturally independent of each other, so assigning different events to different threads is an accepted and viable strategy for implementing coarse-grain parallelism in KF-based tracking software. However, if the goal is to unlock the vector or SIMD potential of modern CPUs or GPUs, this approach is insufficient in itself, since vectorization entails fine-grain parallelism at the level of single instructions. Matrix-based algorithms are generally vectorizable, but only if the rows or columns of the matrices are large enough to fill (say) the vector registers of a CPU; such registers may hold as many as 16 floating-point numbers at a time. Likewise, for the GPU, one looks for operations that can be done in ``warps'' of 32. But the matrices involved in KF-based tracking are 6x6 or smaller and do not meet the basic size requirements.

Fortunately, it is also true that the various tracks created within an event can be built and fitted independently, and this gives us a different possible route for expressing SIMD parallelism in KF-based tracking. Track candidates may be grouped together so that they
propagate in lockstep fashion from one layer of the detector to the next. They may then be extended with compatible hits, where ``compatible'' means that adding a given hit to a track candidate allows the track to pass a $\chi^2$ test. This process results in an updated, and possibly expanded, set of track candidates for propagation to succeeding layers. Of course, vectorization works best when the SIMD operands are contiguous in memory, as this allows the operands to be fetched as already-formed vectors (``coalesced memory accesses'', in GPU terms). As we will see, this preference leads to data structures that are rather different from the ones that would use when Kalman filtering one track at a time. One should now think in terms of groups of small matrices, and how to form SIMD operations from such groups efficiently.

Concurrent processing of track candidates not only enables SIMD processing, it also expands the opportunities for multithreading.  Different threads can now work together 
as they advance groups of track candidates layer by layer. The \mkFit algorithm is structured with three different levels of multithreading: 
\begin{enumerate}
    \item Loop over events: The coarsest level of parallelism occurs over events by processing multiple events concurrently.
    \item Loop over detector regions: Within each event, we divide the seed tracks into regions based on the detector geometry.
    \item Loop over groups of seed tracks: The finest grain parallelism occurs over groups of seed tracks, so that different threads will examine distinct collections of hits on each layer, effectively reducing the memory footprint of each thread.
\end{enumerate}

Overall, our parallelization strategy radically shifts the focus of the KF algorithm. Instead of advancing one track candidate at a time through the detector until all possible tracks from the event have been evaluated and assessed, we consider the global set of track candidates that is present at one layer and propagate it to the next, then to the next, until the final layer is reached. At each layer, current track candidates may be extended by adding a compatible hit, and the retained candidates are again grouped so they can be evaluated via SIMD-style matrix operations as they proceed to the next layer.

\subsubsection{Vectorization using \Matriplex}

\begin{figure}[h]
\centering\includegraphics[width=0.95\linewidth]{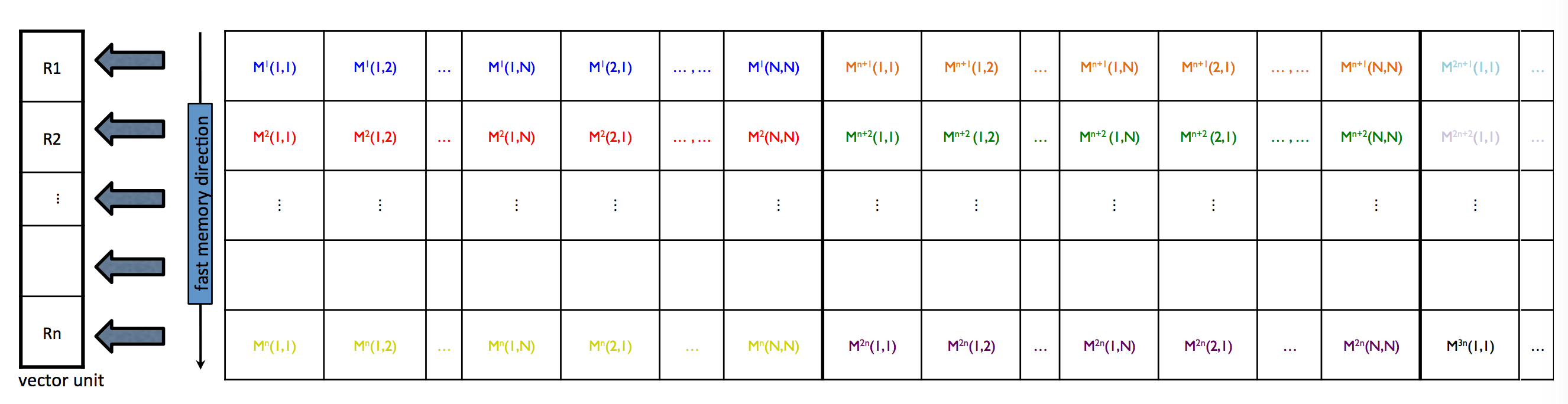}
\caption{Matriplex data structure representation for a matrix size $N$$\times$$N$ and a vector unit size $n$. Due to the use of "matrix-major" storage order, the first elements of $n$ different small matrices $M^i$ are properly aligned for quick SIMD processing. With a single vector instruction, these $n$ elements can be added or multiplied (or both) with their counterparts in $n$ other matrices, vectors, or scalars which are stored similarly.}
\label{fig:matriplex}
\end{figure}

In order to optimize efficient vector operations on small matrices, and to decouple the computational details from the high-level algorithm, we have developed a new matrix library, \Matriplex. The \Matriplex memory layout (Fig.~\ref{fig:matriplex}) uses a matrix-major representation optimized for loading vector registers for SIMD operations on a set of small matrices, using the native vector-unit width on processors with vector units. \Matriplex is similar in concept to other, independently developed solutions~\cite{Kyungjoo}. \Matriplex includes a code generator for defining optimized matrix operations, with support for symmetric matrices and on-the-fly matrix transposition. Patterns of elements that are known by construction to be zero or one can be specified, and the resulting code will be optimized to eliminate unnecessary register loads and arithmetic operations. The generated code can be either standard C++ or macros that map to architecture-specific intrinsic functions.
\Matriplex structures and auto-generated code are used for all KF-related operations on tracks and hits, and in general for all matrix operations. For vectorization of parts of track propagation, where track parameters are transported to a new position using non-linear functions, compiler-assisted vectorization is used.

\subsubsection{Multithreading using TBB tasks}
 
In the preceding sections, we identified several opportunities for multithreading due to the algorithmic independence of events and tracks.  These possibilities include processing in parallel multiple events, detector regions, and groups of seed tracks.  One challenge is that at every level of possible parallelization the workloads are highly irregular: different events have different levels of complexity; different detector regions have variable numbers of layers and hit occupancy, especially in the transition regions where both the barrel and one of the endcaps of the CMS detector must be considered; and different seed tracks will have different numbers of hits, layers, and viable candidates.  Many of these irregularities depend on the path each individual track candidate takes as it traverses the detector, making them inherently unpredictable.

Due to the irregularity of the workloads, static work partitioning schemes proved to be inefficient, largely due to tail effects where threads that happened to be assigned small workloads would wait for threads with larger workloads to complete.  To mitigate the tail effects, we use \Intel Threading Building Blocks (TBB), which allows us to break the workload into relatively fine-grained tasks and balance the workload through dynamic task ``stealing''.  We parallelize at all the levels previously listed, using nested parallel-for loops, which TBB processes efficiently by building a tree of tasks.  For the innermost loop over groups of seed tracks, we use a simple adaptive calculation to attempt to set the task size large enough to minimize overheads but also small enough to allow effective load balancing.  We find that the triply nested structure gives us considerable flexibility to adapt to different workload characteristics, resulting in significant improvements in CPU utilization compared to our initial attempts at static scheduling using OpenMP.

\subsection{Coordinate Representation}

In our implementation of the KF algorithm, we make use of a coordinate representation in global coordinates. This choice is mainly driven by the need to minimize the size of data structures, and in particular those related to the detector description. Working in the local coordinates of each sensor module would require storing the global position and rotation of each module in addition to the local positions of the hits.
In global coordinates, hit locations are simply defined by their 3D positions: $(x,y,z)$.
Helical trajectories of charged particles in a uniform magnetic field are likewise represented using global Cartesian coordinates for the spatial reference point, while using polar coordinates for the momentum vector: $(x,y,z,\phi,\theta,1/p_{T})$, where $\phi$ is the azimuthal angle of the momentum vector, $\theta$ is the angle of the momentum vector with respect to the direction of the magnetic field (assumed to be along the $z$ axis), and $p_{T}$ is the transverse component of the momentum vector with respect to the direction of the magnetic field. 
It is also useful to define an additional variable, pseudorapidity $\eta\equiv-\ln(\tan(\theta/2))$, which will be used in later sections.
While the minimal representation of a helix needs only 5 parameters and ours may look redundant, a sixth parameter is formally needed to define the current position along the helix (path length). 
Our representation also has the advantage that the track and hit positions are immediately comparable.
Covariance matrices for hit and track parameters are defined as 3x3 and 6x6 symmetric matrices, respectively. 
The dimension of these matrices exceeds their rank, so they could be stored in more compact forms; however, defining covariance matrices that directly map the parameter representation is clearly convenient since no transformations are needed. 

\subsection{Detector Description and Navigation}
\label{sec:geom}

The geometry of tracker detectors typically follow a hierarchical structure, from layers in different detector types to the actual sensors on the modules. Tracking algorithms usually navigate through this hierarchy (typically described in terms of templated classes) to locate the actual measurements on the sensors. As described below, \mkFit instead relies on a simplified detector description, with the goal of reducing memory and instruction-level overheads.

\begin{figure}[h]
\centering\includegraphics[width=0.32\linewidth]{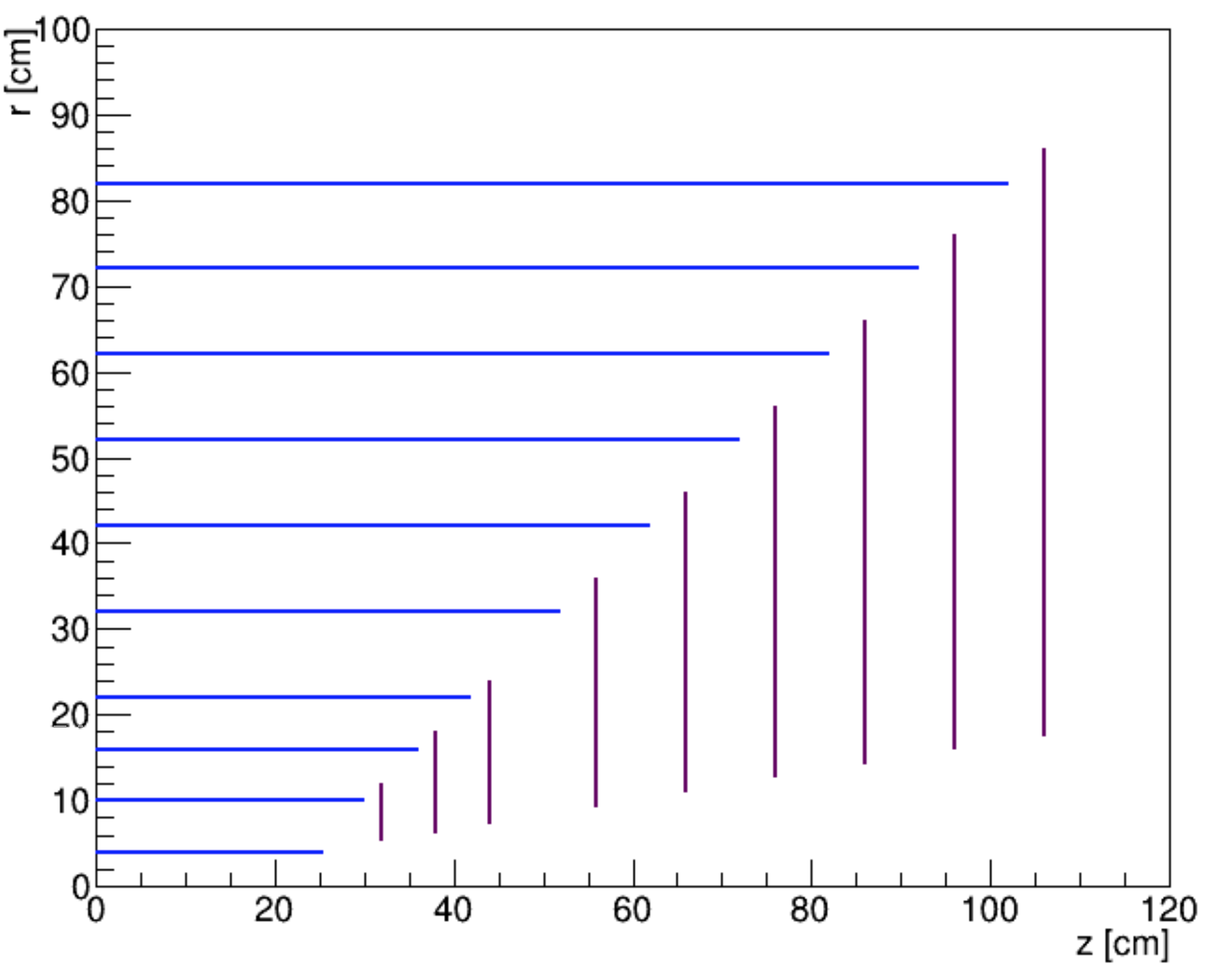}
\centering\includegraphics[width=0.66\linewidth]{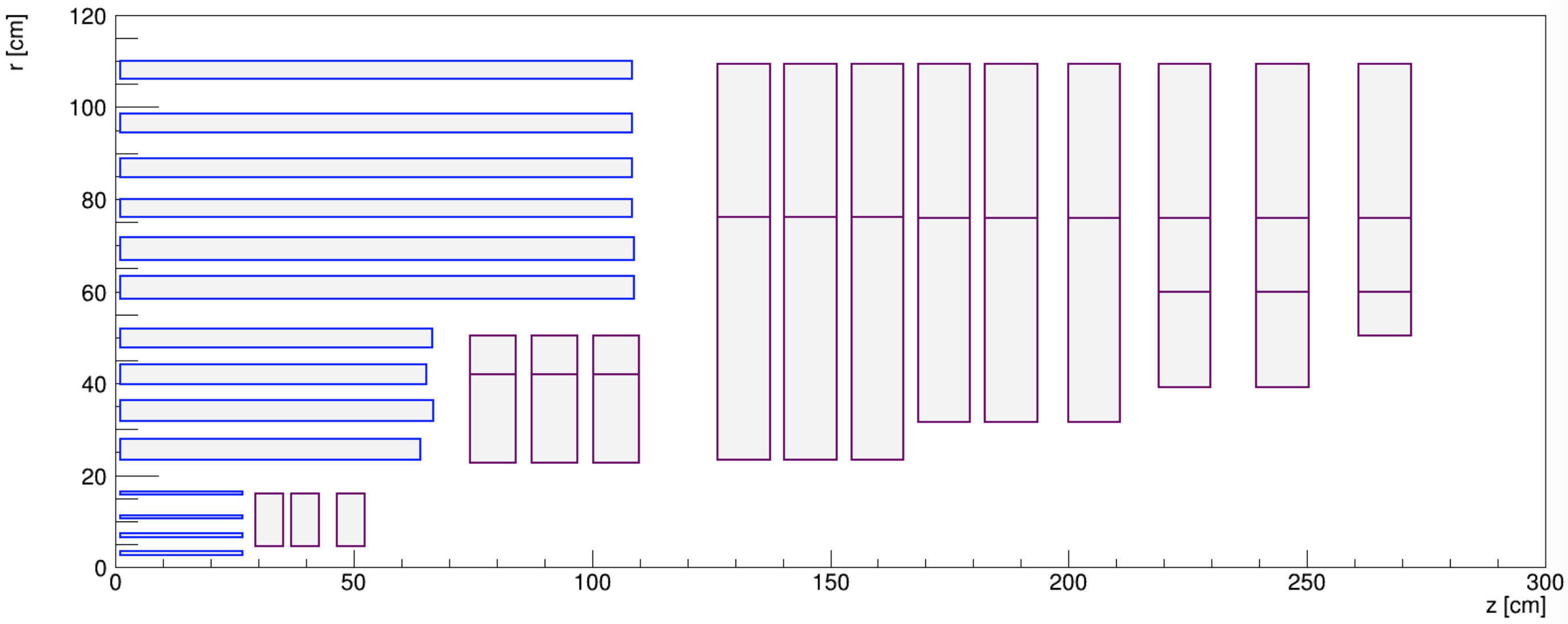}
\caption{Geometry representations as implemented in \mkFit: simple geometry for development (left) and CMS Phase I geometry (right). The rectangles correspond to the geometric $r$-$z$ ranges as defined in the LayerInfo. In order to reduce overheads arising from a full geometry description, \mkFit makes use of a coarse representation of the CMS Phase I geometry; for a description of the actual CMS Phase I geometry see~\cite{CMS:2012sda}.}
\label{fig:geoms}
\end{figure}

The detector geometry and the instructions to navigate through it ("steering parameters") are implemented in \mkFit as a plugin that populates in-memory data structures with the required information. This functionality allows us to support multiple geometry options, which currently include a simple geometry used for development, and the CMS "Phase I" tracker geometry~\cite{CMS:2012sda}; these geometries, as implemented in \mkFit, are shown in Fig.~\ref{fig:geoms}. All detector-specific information resides only in the plugin code. 

Geometry in \mkFit is described as a vector of \emph{LayerInfo} structures that contain the physical dimensions of a layer and parameters and flags relevant for track building. 
Layers are described in terms of inner and outer layer radius and minimum and maximum $z$ coordinate; such a description is sufficient for both barrel (extending along $z$) and endcap layers (extending perpendicular to $z$).
LayerInfo includes information about layer type, 
hit search windows, and an optional hole in the layer coverage; if needed, this structure could be extended for even more general acceptance handling. \mkFit does not implement a description of geometry elements such as individual modules within a layer.

For track building, steering parameters need to be defined for every tracking region. 
Five tracking regions are currently defined based on $\eta$, which is useful for separating the tracks according to the areas of the detector they will encounter. Note that the concept of tracking regions could also be extended to separate tracks according to other kinematic properties of the tracks.
The steering parameters contain, most importantly, a vector of \emph{LayerControl} structures that hold layer indices (mapping into the LayerInfo vector) that need to be traversed during track building. The steering parameters also include layer parameters and flags that are specific for this tracking region, such as tagging layers as possible seeding layers. 
This allows the track building algorithm to be completely agnostic of the detector structure: it simply follows the layer propagation plan in the steering parameters and executes operations in accordance with the control flags in LayerControl and LayerInfo structures.

Track propagation is performed in two steps. In the first step, the track state is propagated to the average radius or $z$ of the layer (for barrel and endcap layers, respectively); here hits within a compatibility window (adjusted for the layer spread with respect to the average coordinate) are identified. In the second step, the track is propagated to the actual position of each candidate hit and a $\chi^2$ is computed to determine which hits match the track best. In the version of the algorithm used for this paper, \mkFit can only add one hit per layer to each track candidate. If a track crosses two overlapping modules on the same layer, competing track candidates may be created, but both hits cannot be stored in the same candidate. The ability to collect more than one hit per layer is currently in development, but as we will see in Sec.~\ref{sec:Results}, the physics performance is already sufficient for the purpose of the present demonstration.

For the CMS Phase I geometry, we include the effects of multiple scattering and energy loss by defining two-dimensional arrays, indexed in $r$-$z$, with values of the radiation length and the material composition term in the Bethe-Block formula~\cite{Tanabashi:2018oca,CMS-PAS-TRK-10-003,Sirunyan:2018icq}. These constants account for the amount of material a particle would have to traverse in propagating from module to module. 
\mkFit supports the usage of both constant and parameterized magnetic fields, and the type of field description can be selected each time propagation is required in the code.

\subsection{Branching of Candidates}

As already mentioned, track building is a combinatorial process: when a track candidate is propagated to a layer, hits located within a compatibility window are probed and those that yield the best $\chi^2$ lead to new candidates (branching) that are then propagated to the next layer. The number of allowed candidates branching off a seed is limited by a configuration parameter, MaxCand. The combinatorial behaviour leads to several computational challenges.

The first challenge is data locality. The key to reducing this problem is to spatially partition hits and tracks so that hits are accessed efficiently, and so that tracks, when processed concurrently, access overlapping sets of hits. As mentioned, tracks are partitioned into tracking regions that define the general sequence of layers the tracks can cross. Within each tracking region, seeds are sorted by $\eta$ (based on the outermost hit of the seed), so tracks are processed in this order on every layer, and hits too are typically accessed in the same $\eta$ order. Additional second-order sorting of tracks in $\varphi$ was considered but has not been implemented, because of the expected divergence of tracks in the $\varphi$ space as they curve differently in the magnetic field. 
In order to access hits efficiently, it is helpful to introduce a new coordinate variable $q$ that stands for $z$ in the barrel and for $r$ in the endcaps; this allows for a common description of \mkFit algorithms in both detector regions. Indices of hits in every layer are split into $q$-bins (which are equivalent to $\eta$-bins) and each of them is sorted on $\varphi$. Algorithmically, a single sorting value is calculated for each hit and all hit indices for a given layer get sorted with a single radix sort operation. Within each $q$-bin, an equidistant partitioning table in $\varphi$ is created to provide a fast look-up of compatible hits.

The second challenge is that different seeds may produce a different number of candidates, and this may lead to a load imbalance in the concurrent processing. In order to minimize this problem, at each layer \Matriplex objects are populated by candidates from different seeds so that all SIMD units are maximally occupied.

Finally, each new branch requires creating a copy of the original candidate, and this operation is serial. To mitigate the impact from this serial work, we moved copying outside of all vectorizable operations into what we term the ``clone engine". The clone engine approach only copies the best $N\leq$ MaxCand track candidates per seed after reading and sorting a bookkeeping list of all compatible hits. This is in contrast to a first attempt at the combinatorial KF, which copied a candidate each time a hit was deemed compatible, then sorted and kept only the best $N$ candidates per seed after all the possible hits on a given layer for all the input candidates were explored. The clone engine leads to a speedup of about 20\%.

\subsection{Duplicate removal steps}
\label{sec:dr}

Before passing seeds to \mkFit for track building, the seed collection is ``cleaned" by removing multiple instances of seeds that are most likely based on hits belonging to the same outgoing particle. The cleaning algorithm compares the fitted seed parameters $p_\mathrm{T}$ , $\eta$, and $\phi$ to eliminate duplicate seeds and is tuned so as to not cause any drop in track building efficiency for high pileup events. In the case of the CMS Phase I geometry, the duplicate seeds arise due to detector module overlaps that are rather significant, especially in the endcaps.
Low $p_\mathrm{T}$ tracks (below 2 GeV/c) are more affected due to bending in the $r$-$\phi$ plane during their flight through the detector. Multiplicity of seeds from a single particle frequently reaches 8 and can go as high as several tens. To further reduce the duplicate rate in \mkFit, a second duplicate removal step occurs after the track building. This step first compares the $p_\mathrm{T}$, $\eta$, and $\phi$ parameters of the built tracks and then checks how many hits are shared between the tracks. These dedicated duplicate removal steps are not necessary for the standard track building algorithm used by CMSSW; CMSSW processes seeds sequentially, and a seed is rejected if all of its hits have already been used by a track candidate found earlier.

\subsection{GPU implementation}

Given the recent trend in the computing market strongly favoring GPUs, in the longer term we plan to work on a GPU implementation of \mkFit. While a large fraction of the work we have done so far should apply just as well to GPUs (which also rely on two types of parallelization, over data and over tasks), large differences are introduced by the different memory hierarchy and the need for data transfers from the host. 
These are significant issues for an algorithm with low arithmetic intensity such as track building.
A separate challenge is that a GPU-only implementation would require a duplication of code and maintenance; to solve this challenge, we plan to explore possible portable solutions to allow the same code to run on both GPUs and multicore CPUs. All combined, an efficient GPU implementation of \mkFit is not a trivial task and may not necessarily outperform the current version, or at least not without substantial effort.



\section{Results}
\label{sec:Results}

\subsection{Introduction}

In this section, we demonstrate the performance of the \mkFit algorithm. First, we present results for physics quantities of interest such as efficiency, fake rate, and duplicate rate. Second, we discuss the computational performance, including timing comparisons with the nominal CMSSW tracking algorithm. The input data for these tests is a simulated data sample of \ttbar events with an average pileup per event of 50 using the Phase I CMS geometry in 2018 and assuming realistic detector conditions. Studies with higher average pileup values are also planned but require updating the CMS geometry to incorporate the planned tracker upgrades \cite{PhaseIIpaper}. There are two setups that can be used to validate the performance of the \mkFit algorithm: standalone and integrated. The standalone setup  (Sec.~\ref{physics} - \ref{computing}) does not require the CMSSW environment and is used for development and controlled tests of the algorithm. It includes a suite of computational benchmarks that are useful for identifying potential gains or losses in computational performance and for testing proposed changes and improvements to \mkFit. External profiling tools, such as those provided by \Intel and \texttt{TAU}~\cite{TAU}, were also used for more fine-grained inspection of potential bottlenecks in source code.
The integrated setup (Sec.~\ref{CMSSW}) runs \mkFit within CMSSW, which gives us access to the standard CMS validation tools as an additional check on the performance of \mkFit. 

When processing CMS events, \mkFit relies on hit and seed data to be provided externally. In the standalone setup, \mkFit reads the input hit and seed data from a binary file created by a converter application. Additionally, the binary file can also contain vectors of simulated tracks and reconstructed tracks as found by standard CMS tracking used in the validation of \mkFit's performance. In the integrated setup, dedicated CMSSW modules convert the input hits and seeds into the \mkFit format. 

\subsection{Platforms and General Configuration}

The primary architectures used in the testing and development of the \mkFit algorithm are \Intel multicore and many-core devices, i.e., \Intel \Xeon processors and \Intel \XeonPhi processors and coprocessors, although the latter product line has been effectively terminated. 
We use the following machines for presenting physics and
computational performance:
\begin{itemize}
    \item ``KNL'' (Knights Landing) -- 64 cores: \Intel \XeonPhi processor 7210 {@} 1.30\unit{GHz}
    \item ``SKL-SP'' (Skylake-Scalable Processor) -- dual socket x 16 cores: \Intel \Xeon Gold 6130 processor {@} 2.10\unit{GHz}
\end{itemize}

In order to exploit the two AVX-512 vector processing units per core featured on these machines, \mkFit is compiled using \Intel \cpp Compiler version 19.0.4 with the instruction set native to the device. 
Intrinsic functions for memory alignment and access are used whenever possible. Additional flags are included in the compilation to perform routine code optimizations (e.g., function inlining and branch prediction from speculative execution) and to ensure high utilization of vector registers. We enable \Intel Hyper-Threading Technology on all machines, and test up to the maximum number of hardware threads in each machine, i.e., 256 for KNL and 64 for SKL-SP.

The machines were configured in the following way for the tests below. The clock frequency scaling governor was set to ``performance'' mode, which signals to the clock frequency manager to prevent ramping down the clock frequency unless absolutely necessary. \Intel Turbo Boost was disabled to reduce confusion when measuring the computational scaling behavior as a function of the number of threads. 

\subsection{Physics Results}
\label{physics}

In this section we compare the physics performance of \mkFit to that of the track building algorithm used in CMSSW version 10\_4\_0\_patch1 in terms of metrics that probe the correctness and completeness of the track hits identified during track building. We focus on the first tracking iteration of the offline CMS tracking, seeded by quadruplets of hits located in the pixel detector~\cite{Pantaleo:2293435}.

\begin{figure}[!htb]
  \centering
  \includegraphics[width=0.48\linewidth]{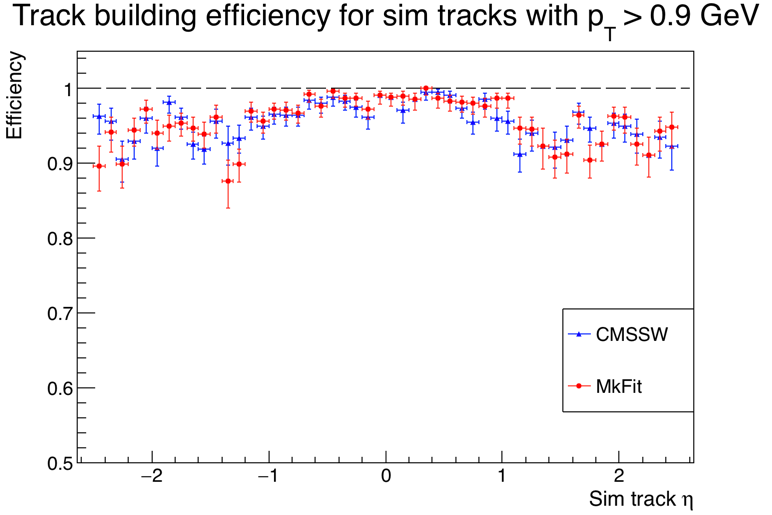}
  \includegraphics[width=0.48\linewidth]{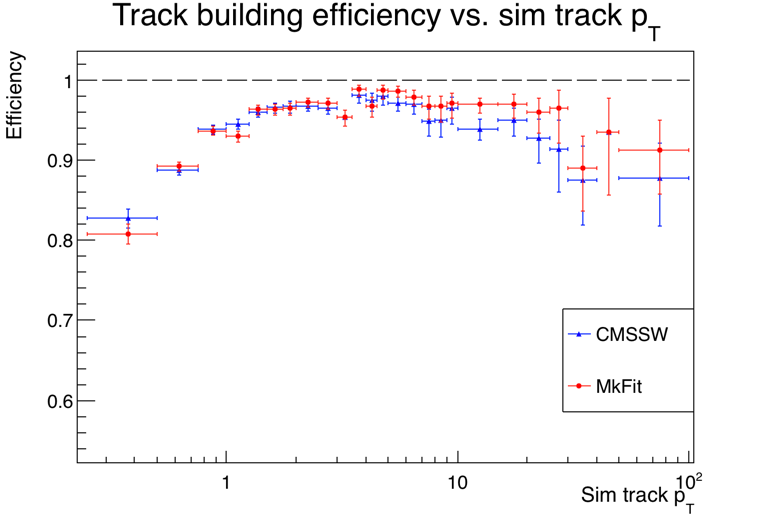}
  \caption{Efficiency of the \mkFit (red) and nominal CMSSW (blue) track building algorithms as a function of the track $\eta$ (left) and $p_\mathrm{T}$ (right). The efficiency with respect to track $\eta$ is calculated for tracks with $p_\mathrm{T} > 0.9$ GeV. The efficiency is defined as the fraction of simulated tracks that are matched to at least one reconstructed track; only simulated tracks matched to a seed are considered. Sample used: first CMS offline tracking iteration for \ttbar events with $<$PU$>$=50 and CMSSW version 10\_4\_0\_patch1.}
    \label{fig:eff}
\end{figure}

The efficiency of the \mkFit algorithm is evaluated with respect to simulated tracks that are prompt (that is, originate at or near the collision point), within the strip detector acceptance ($|\eta|<2.5$), and matched to a track seed. The last requirement is applied in order to factor out the efficiency for finding seeds, which are an external input to \mkFit. A reconstructed track is considered matched to a simulated track if more than 75\% of the reconstructed track hits are shared, including the hits from the seed. The track building efficiency is defined as the fraction of simulated tracks that are matched to at least one reconstructed track. Figure~\ref{fig:eff} shows the efficiency as a function of $\eta$ for tracks with $p_\mathrm{T} > 0.9$ GeV and the efficiency as a function of track $p_\mathrm{T}$. The \mkFit algorithm is at least as efficient as the nominal CMSSW algorithm for all values of track $\eta$ and $p_\mathrm{T}$.

\begin{figure}[!htb]
  \centering
  \includegraphics[width=0.48\linewidth]{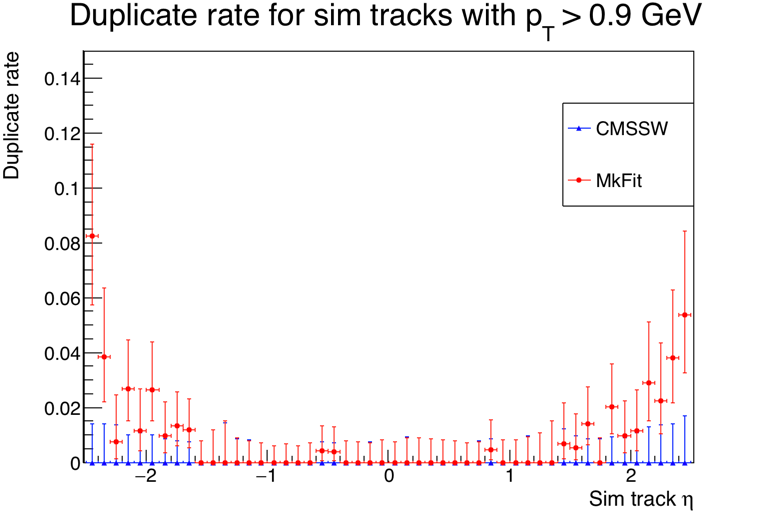}
  \includegraphics[width=0.48\linewidth]{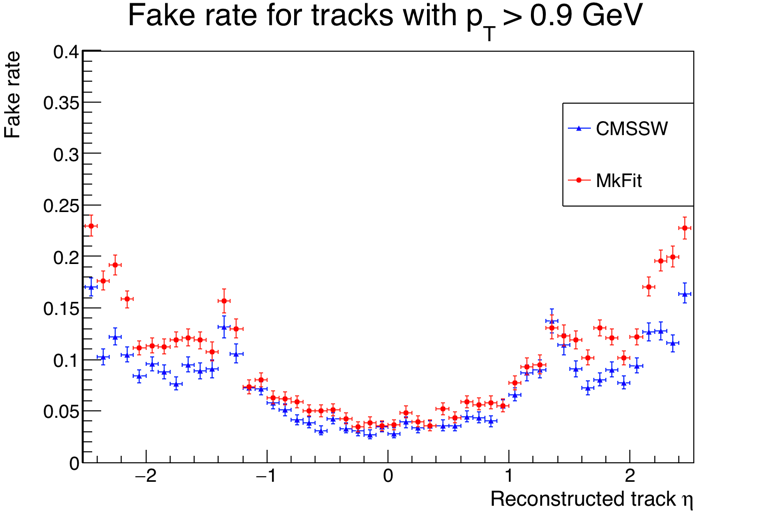}
  \caption{Duplicate rate (left) and fake rate (right) of the \mkFit (red) and nominal CMSSW (blue) track building algorithms as a function of the track $\eta$, for tracks with $p_\mathrm{T} > 0.9$ GeV. The duplicate rate is defined as the fraction of simulated tracks matched to at least one reconstructed track that are matched to multiple reconstructed tracks. The CMSSW values are nearly zero. The fake rate is defined as the fraction of reconstructed tracks that are not matched to a simulated track. Sample used: first CMS offline tracking iteration for \ttbar events with $<$PU$>$=50 and CMSSW version 10\_4\_0\_patch1.}
  \label{fig:dr}
\end{figure}

Figure~\ref{fig:dr} (left) shows the duplicate rate as a function of $\eta$ for tracks with $p_\mathrm{T} > 0.9$ GeV.
 The denominator of the duplicate rate includes all simulated tracks that are matched to at least one reconstructed track, and the numerator includes all simulated tracks that are matched to multiple reconstructed tracks.
 The duplicate rate is worse for higher values of $\eta$ due to the larger number of overlapping detector modules leading to multiple track seeds for a single charged particle. The two duplicate removal steps outlined in Sec.~\ref{sec:dr} reduce the overall duplicate rate in \mkFit from over 30\% to less than 1\%.
 
The fake rate is defined as the fraction of reconstructed tracks that are not matched to a simulated track. The fake rate of the \mkFit algorithm is shown in Figure~\ref{fig:dr} (right) as a function of $\eta$ for tracks with $p_\mathrm{T} > 0.9$ GeV; the performance of \mkFit and CMSSW in terms of fake rate is similar, with \mkFit 3.5\% higher in overall absolute terms.
Note that the results shown in Figure~\ref{fig:dr} include all tracks produced by the track building step; the fake rate is further reduced for both \mkFit and CMSSW tracks by additional track selections that are applied after the final fit in CMSSW. 

\begin{figure}[!htb]
  \centering
  \includegraphics[width=0.49\linewidth]{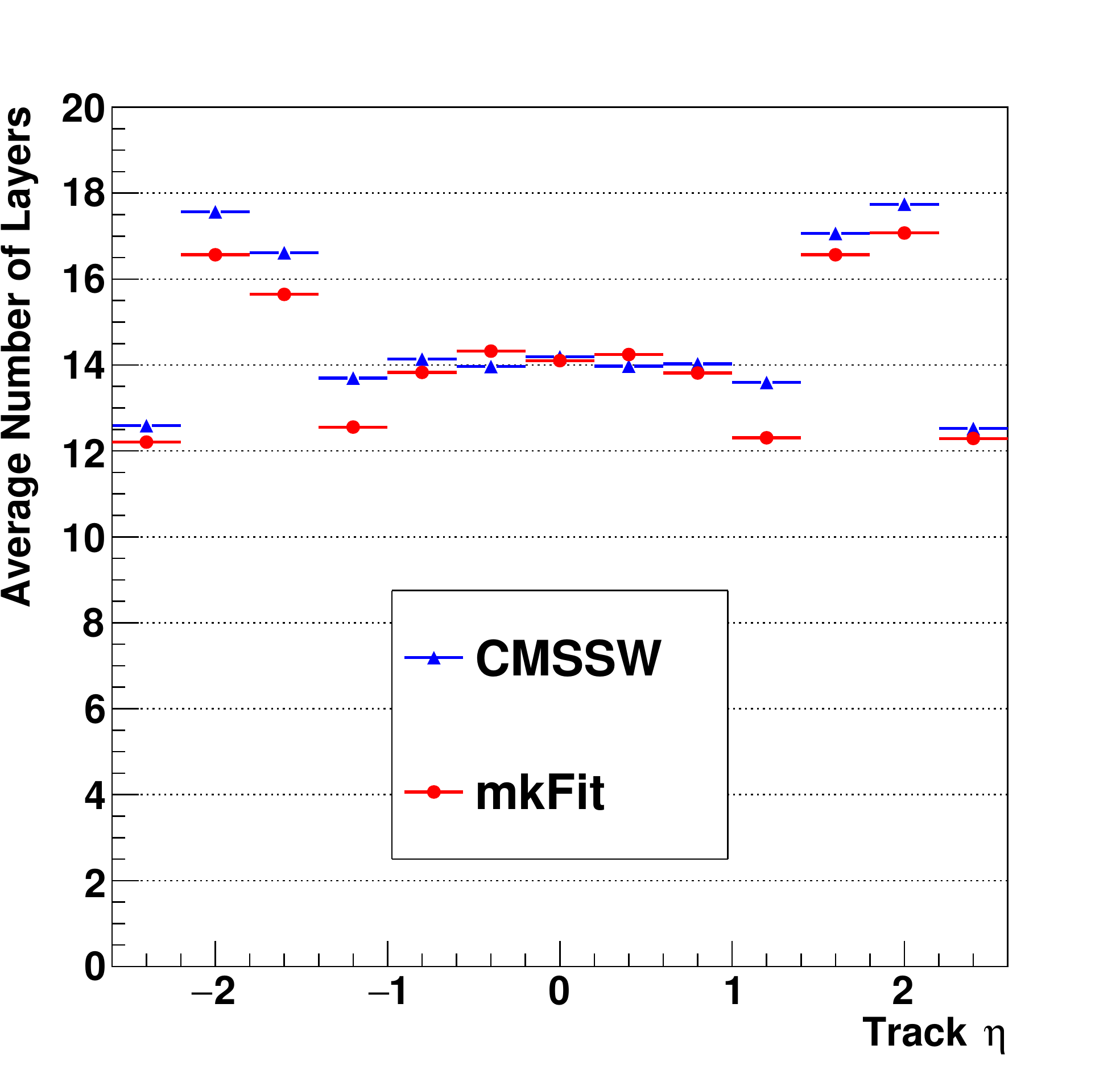}
  \caption{Number of layers with found hits in tracks reconstructed by the \mkFit algorithm (red) and tracks reconstructed by CMSSW (blue) as a function of $\eta$. The layer count includes the seed. Tracks are required to be matched to a simulated track. Sample used: first CMS offline tracking iteration for \ttbar events with $<$PU$>$=50 and CMSSW version 10\_4\_0\_patch1.}
  \label{fig:nLayers}
\end{figure}

It is also important to ensure that the tracks reconstructed by \mkFit are similar in quality as the tracks reconstructed by CMSSW, e.g., by verifying that the algorithm collects the majority of hits along the track. Figure~\ref{fig:nLayers} shows the number of layers with found hits in tracks that are reconstructed by \mkFit and CMSSW; the overall average number of layers is 15.1 for CMSSW and 14.9 for \mkFit.

In summary, while not identical, the \mkFit results in terms of the metrics presented here are sufficiently close to those of the nominal CMSSW algorithm so that, for the purpose of the present demonstration, the two algorithms can be considered equivalent from the physics performance point of view.

\subsection{Computing Results}
\label{computing}
This section outlines the computational performance of \mkFit, measured primarily with a set of benchmarks 
that test the scaling behavior of \mkFit as a function of increased resources. 
We first measure the vectorization and multithreading performance solely of the track building subroutine within \mkFit, ignoring the time for I/O, seed preparation, hit organization, etc. In both tests, we process a total of 100 events and sum the build times for all but the first event. In these tests, events are processed one at a time, i.e., the number of concurrent events is one. We also limit the vectorization test to a single thread, to factorize speedups due to vectorization from those due to multithreaded parallelism. 
The vectorization test is performed by varying the vector width of \mplex in multiples of two, effectively limiting the use of vector registers to the number of floating point numbers in the width of \mplex; otherwise, the compiler is instructed to auto-vectorize and optimize the code as much as possible. It is worth noting that when disabling auto-vectorization with \mplex width set to one, the track building time slows by a further 15\%, thus demonstrating that part of the code is still vectorized when \mplex width is set to one.
The multithreading test measures the performance on top of the maximum vectorization performance by gradually increasing the number of threads. 

\begin{figure}[htbp]
  \begin{center}
    \includegraphics[width=0.49\columnwidth]{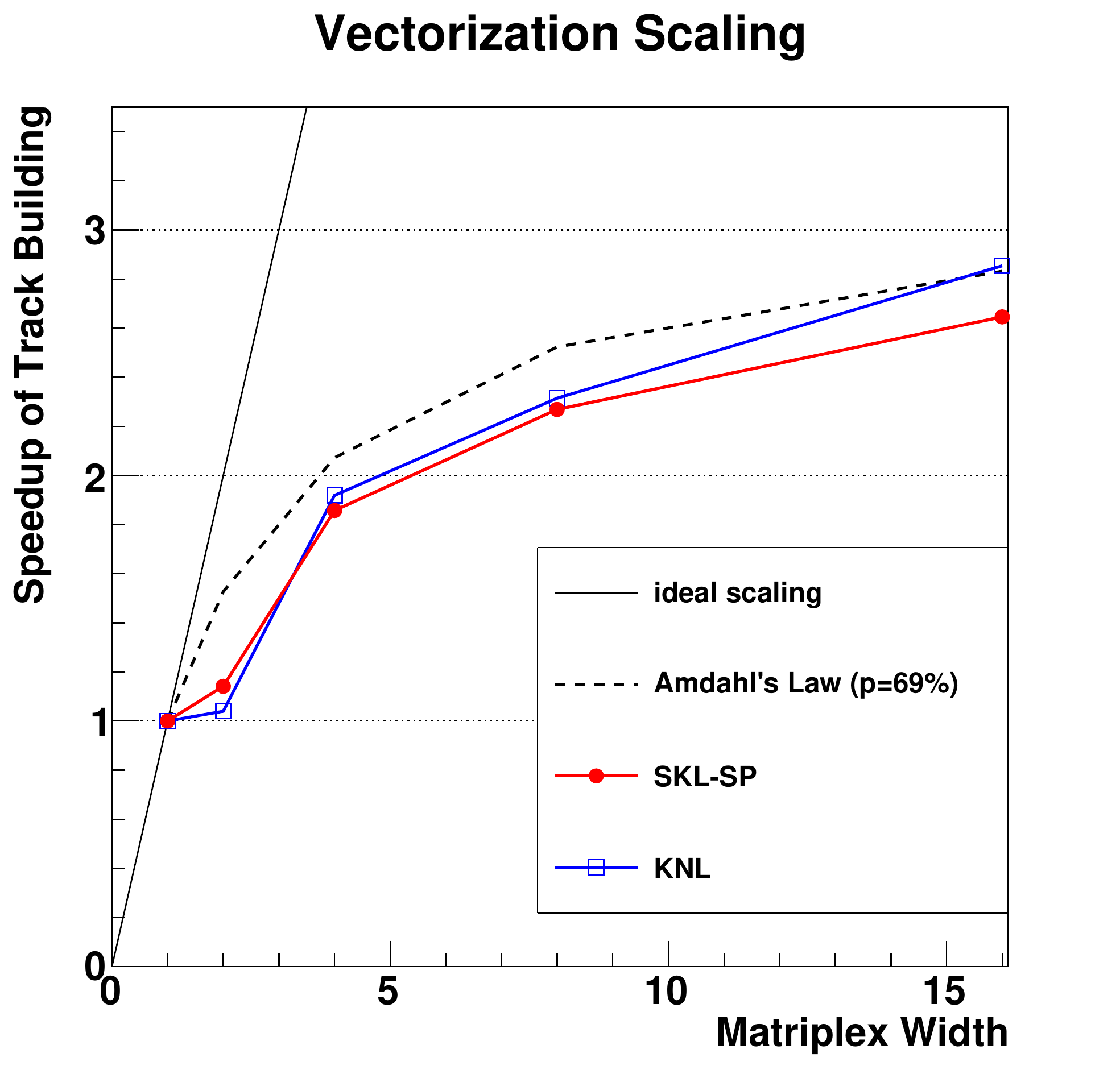}
    \includegraphics[width=0.49\columnwidth]{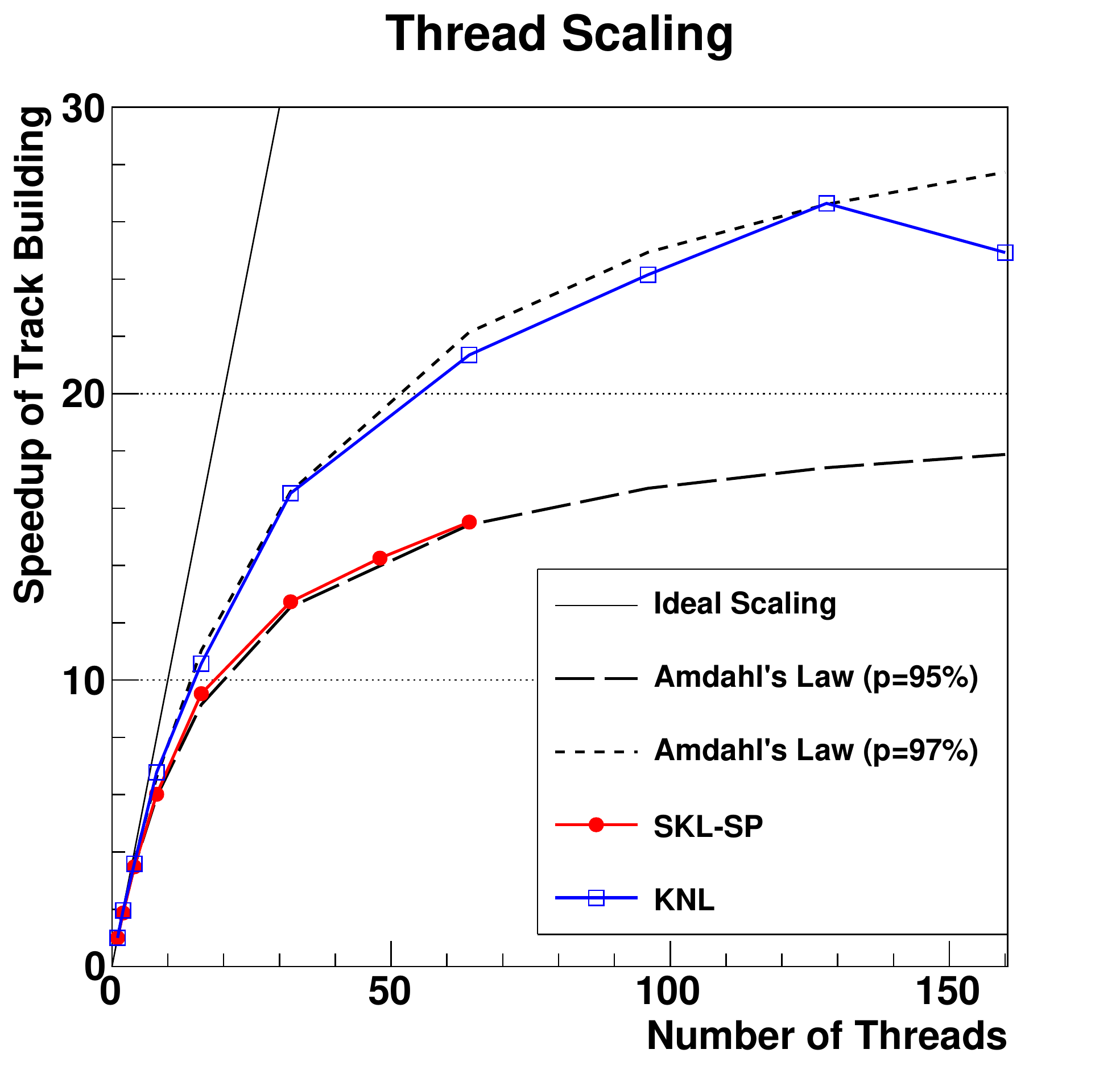}
    \caption{Vectorization speedup as a function of the \mplex width (left), and multithreading speedup as a function of the number of threads allotted (right), for \mkFit track building (only) on KNL, and SKL-SP. The code is complied with the instruction set native to the processor in each case, regardless of \mplex width. The ideal speedup curve (solid line) assumes perfect scaling as a function of increased resources; speedup curves based on Amdahl's Law (dotted and dashed lines) are shown for comparison.}
    \label{fig:mkfit_time_benchmarks}
  \end{center}
\end{figure}

To demonstrate the scaling behavior as a function of additional resources, we compute the speedup: for each point tested, we divide the wall-clock time measured for vector width (number of threads) equal to one by the wall-clock time measured for vector width (number of threads) equal to $N$. The speedup results of the vectorization and multithreading tests are shown in Fig.~\ref{fig:mkfit_time_benchmarks}.

The scaling results in Fig.~\ref{fig:mkfit_time_benchmarks} can be interpreted using Amdahl's Law~\cite{ref:amdahl}, which relates the speedup due to increased resources (from either vectorization or multithreading) to $p$, the fraction of the program that is parallelizable. The serial fraction ($1 - p$) necessarily limits the speedup, since additional resources cannot increase the performance of the serial portion.
We infer $p$ from the results by writing Amdahl's Law in the form
\begin{equation}
  \label{eq:amdahl}
  p = \frac{1-(1/S)}{1-(1/R)},
\end{equation}
where $S$ is the measured speedup, and $R$ is the ratio of available to original resources. 
Given measured speedups on SKL-SP (KNL) with a maximum of 2.7 (2.9) at a \mplex vector width of 16, the fraction of the track building code that is effectively vectorized is nearly 70\%. Notice that the speedup when shifting from \mplex width of one to width of two is consistently smaller than for larger widths, possibly because there are no native instructions for a vector size of two.
Multithreading tests show maximum speedup values of $S=15.5$ ($S=26.6$) at 64 (128) threads on SKL-SP (KNL), thus implying that the fraction of code that is parallelized is at least 95\%.

\begin{figure}[htbp]
  \begin{center}
    \includegraphics[width=0.48\columnwidth]{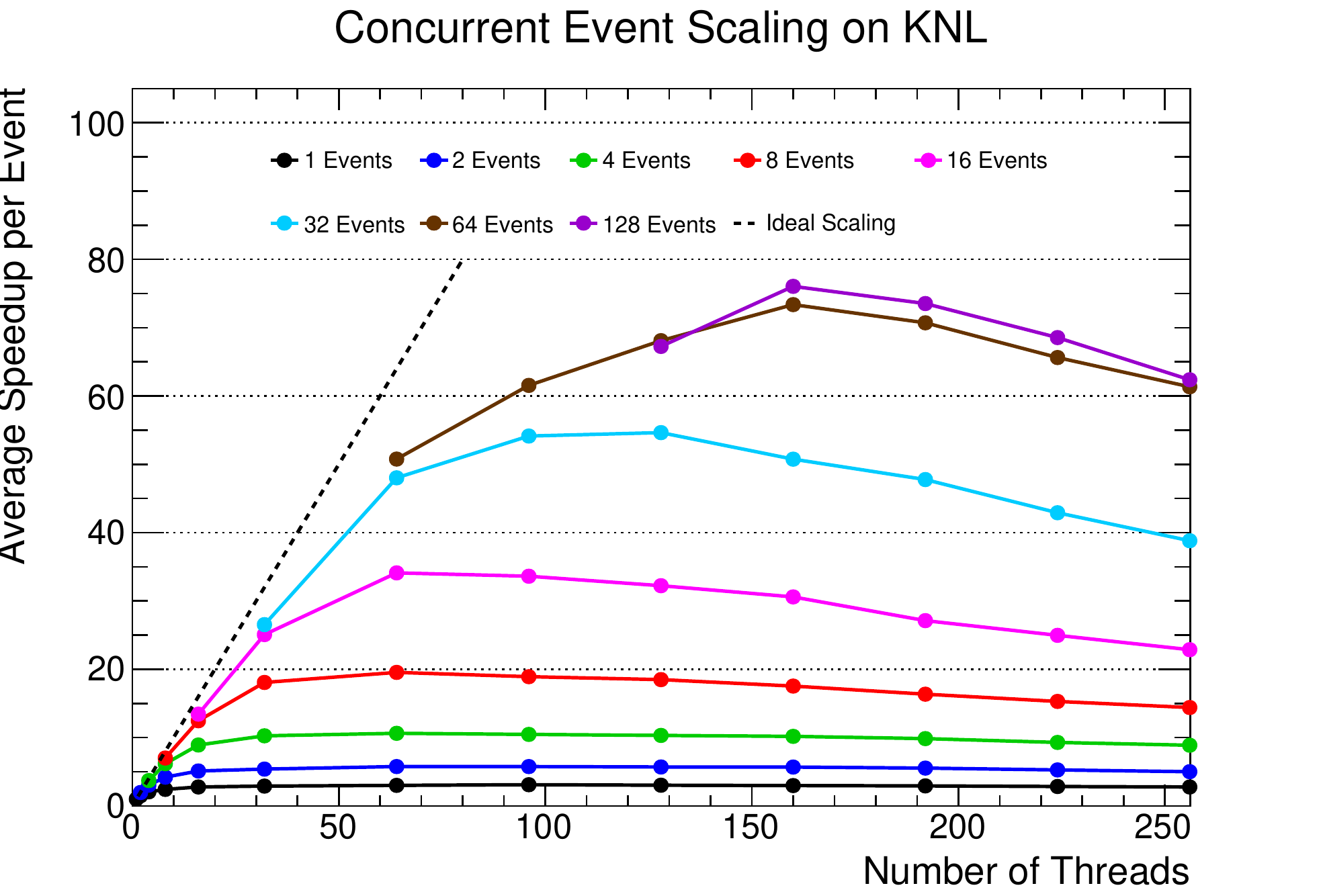}
    \includegraphics[width=0.48\columnwidth]{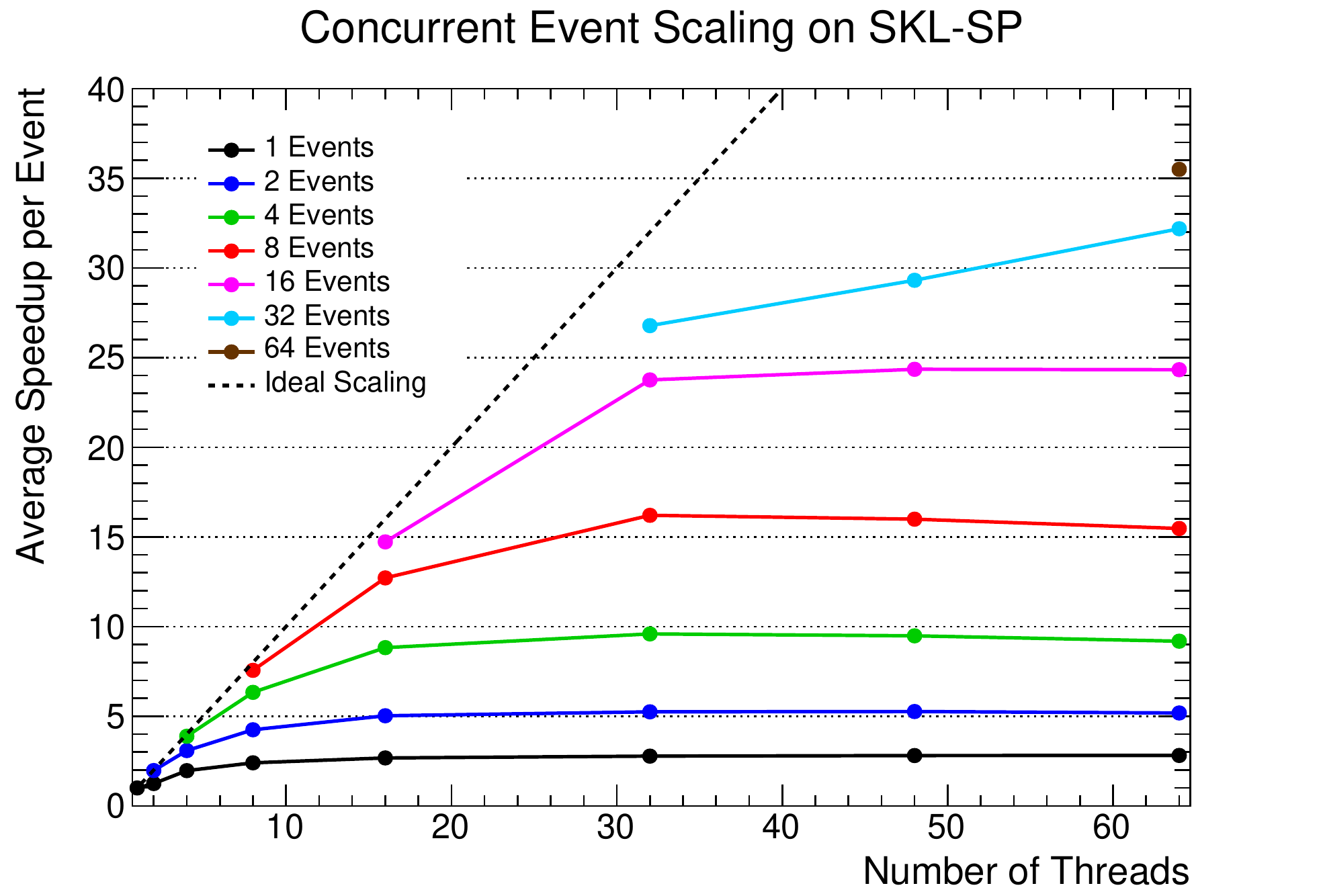}
    \caption{Speedups with respect to one thread and one concurrent event, when varying the number 
      of threads for a fixed number of concurrent events, on KNL (left) and SKL-SP
      (right). Speedups are based on the full \mkFit loop time, which includes
      I/O and seed preparation in addition to track building.}
    \label{fig:mkfit_time_meif}
  \end{center}
\end{figure}

We can also measure the speedup achieved by processing multiple events concurrently. In this case, the time used for computing the speedup is the average wall-clock time per event for processing \mkFit, which now includes sections outside of the track building algorithm. In order to obtain full thread utilization, we increase the workload to reconstruct 5000 events times the number of events processed concurrently. The speedup results with respect to one thread and one concurrent event are displayed in Fig.~\ref{fig:mkfit_time_meif}. As can be seen, by processing events concurrently, latencies between events are hidden and a very impressive scaling can be achieved. In fact, we measure a maximum speedup of 35.2 (76.0) on SKL-SP (KNL), appreciably larger than the total number of physical cores.

\begin{figure}[htbp]
  \begin{center}
    \includegraphics[width=0.48\columnwidth]{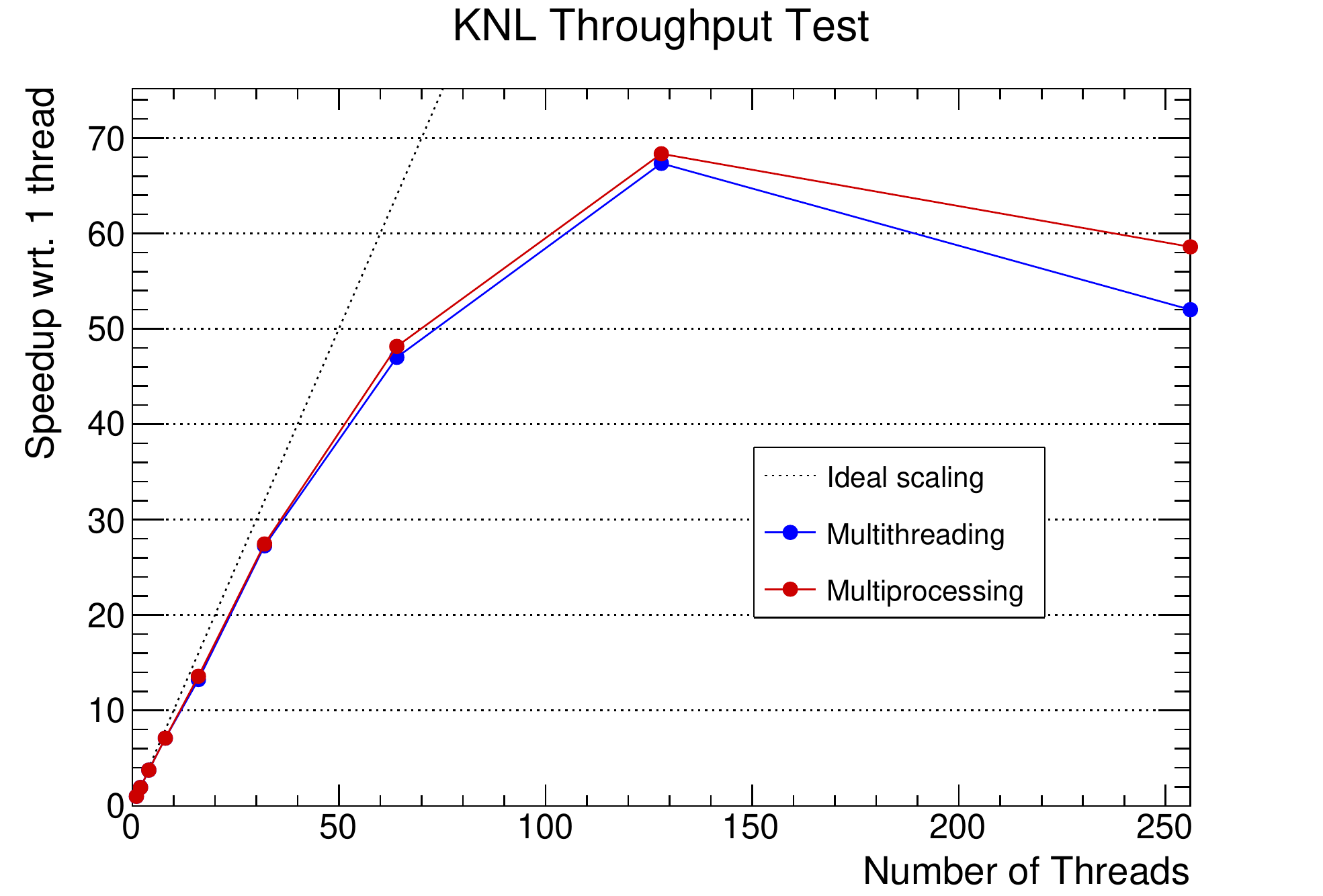}    
    \includegraphics[width=0.48\columnwidth]{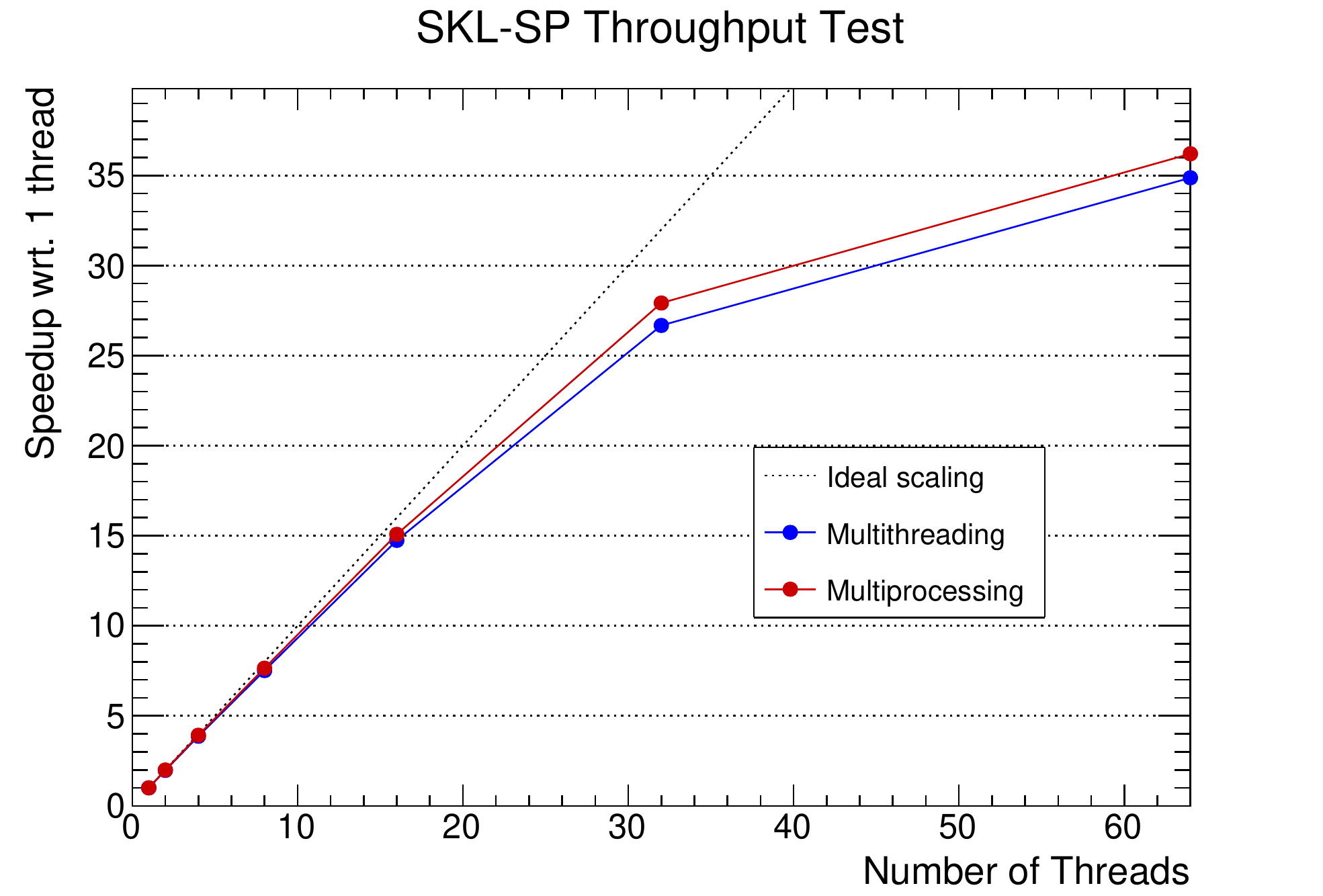}
    \caption{Speedup as a function of the number of threads for the full \mkFit loop time, 
      when the number of threads in one instance equals the number of concurrent events (blue), and the number of 
      single-threaded instances equals the number of threads (red), for KNL (left) and SKL-SP
      (right).}
    \label{fig:mkfit_time_thru}
  \end{center}
\end{figure}

With a similar configuration, we can also measure the total throughput in events per second for \mkFit, comparing the performance in a multithreaded instance to that when multiprocessing a set of instances. In this context, multiprocessing refers to launching multiple instances of \mkFit each with a single thread, whereas multithreading refers to a single instance of the \mkFit program that is launched with multiple threads. For ease of comparison, we set the total number of concurrent events equal to the number of threads in the multithreaded throughput test, and then use that as the number of instances for the corresponding multiprocessing test.
The results of these tests are shown in Fig.~\ref{fig:mkfit_time_thru}.  On a single thread, the throughput on the SKL-SP is 3.4 times higher than the throughput on the KNL. With fully loaded machines, the difference reduces to a factor of 1.8. The points along the multithreaded curve can be mapped to the first point along each of the multiple event-in-flight curves in Fig.~\ref{fig:mkfit_time_meif}, where the total number of concurrent events equals the number of threads for those tests.
The throughput tests demonstrate that the multithreaded \mkFit produces about 3.5\% (1.5\%) less throughput compared to the multiprocessed \mkFit at maximum load on the SKL-SP (KNL) machine. While multiprocessing may achieve an ever-so-slight advantage in throughput, it requires many instances of a program, where every instance and its data must be loaded into memory. An efficient multithreaded program can share resources, and therefore significantly reduce the memory footprint. This  comes at the expense of some overhead to the operating system scheduler that must manage scheduling tasks and dividing resources. The fact that the overhead is proven to be at the precent level demonstrates the excellent division of tasks within \mkFit. 

\subsection{Integration in CMSSW}
\label{CMSSW}

The \mkFit algorithm is included in the CMS software distribution as an external package, along with dedicated processing modules within CMSSW. There are separate modules for packaging the input data (i.e., seeds and hits) into the format expected by \mkFit, for executing the \mkFit algorithm, and for reorganizing the results from \mkFit back into the nominal CMS data formats. These modules provide high-level configuration and steering of the \mkFit execution, and they can be used as a drop-in replacement for the default pattern recognition module in the CMS tracking. The data format conversions between CMSSW and \mkFit introduce a sizeable overhead to the execution time, up to about about 25\% of the \mkFit time; in the future such overhead could be mitigated or removed by harmonizing the data format definitions. This model of inclusion allows the \mkFit code to remain independent of CMS particularities, while also allowing \mkFit to be developed and tested in a more lightweight environment compared to the full CMSSW. 

Within this setup we can directly compare the time performance of the first iteration of offline track building when performed with \mkFit versus CMSSW. 
This test uses just a single thread on SKL-SP, processing 1000 events. 
The \mkFit algorithm is compiled with the \Intel \cpp Compiler (version 19.0.4), and
CMSSW version 10\_4\_0\_patch1 is compiled with the GNU Compiler Collection (version 7.3.1) as released by CMS.

\begin{figure}[htbp]
  \begin{center}
    \includegraphics[width=0.6\columnwidth]{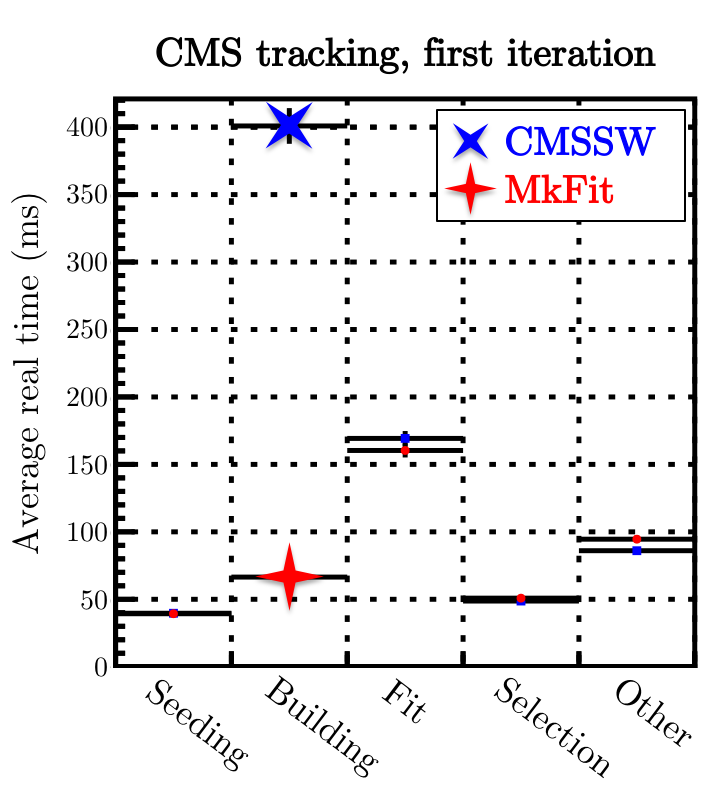}
    \caption{Comparison of the single-thread time to run \mkFit as an external within the CMSSW 
      framework (red) to the time to run the standard first iteration of offline tracking in CMSSW (blue). Note that \mkFit is used as a replacement for the building step only. \mkFit 
      achieves a speedup $>$6x over CMSSW in the track building stage. Time measured on SKL-SP using \ttbar events with $<$PU$>$=50 and CMSSW version 10\_4\_0\_patch1.}
    \label{fig:mkfit_vs_cmssw}
  \end{center}
\end{figure}

The results of this test can be seen in Fig.~\ref{fig:mkfit_vs_cmssw}. Note that \mkFit is used as a replacement for the track building only. With further development, the \mkFit approach could be adapted for the track fitting in addition to the track building, but that is not part of the present work.
\mkFit achieves a $>$6x speedup over CMSSW for the track building. 
The measured times include all overheads, in particular the data format conversions mentioned earlier. 
The 6x speedup includes speedups from vectorization (as shown in Fig.~\ref{fig:mkfit_time_benchmarks}), in addition to 
general algorithmic improvements, such as the 
lightweight representation of the detector geometry that was described in Sec.~\ref{sec:geom}.
Notably, the CMSSW track fitting now takes longer than \mkFit track building. 
In other words, using \mkFit, the track building is no longer the most time-consuming step in CMSSW reconstruction.

\section{Conclusions and Outlook}
\label{sec:outlook}

The KF-based track building algorithm is the main driver in CPU time increase at the LHC with increasing instantaneous luminosity. We successfully re-engineered the KF-based track building algorithm for parallel processing: about 70\% of the core algorithm is effectively vectorized (speedup close to 3x), and multithreading achieves speedups exceeding the number of available physical cores with scaling close to the multiprocessing limit. The physics performance of the re-designed algorithm is comparable to state-of-the-art algorithms, with further fine-tuning still possible. Tests within the CMS reconstruction framework show that \mkFit is faster than the default algorithm on the offline first tracking iteration, and that track building is now faster than track fitting. These results demonstrate that \mkFit is a viable solution to the timing problem of charged particle tracking at the LHC. Work is underway towards a full integration in the CMS experiment, including application to multiple tracking iterations and integration in the HLT configuration, and, in the longer term, towards a GPU-friendly implementation.

\section*{Acknowledgments}

This work is supported by the U.S. National Science Foundation, under the grants PHY-1520969, PHY-1521042, PHY-1520942, PHY-1624356, and OAC-1836650, and by the U.S. Department of Energy, Office of Science, Office of Advanced Scientific Computing Research and Office of High Energy Physics, Scientific Discovery through Advanced Computing (SciDAC) program. We are thankful to the CMS Collaboration for providing access to the CMSSW software framework and the configurations to produce the simulated samples used for testing.


\providecommand{\href}[2]{#2}\begingroup\raggedright\endgroup

\end{document}